\documentclass[a4paper,11pt]{article}
\usepackage{pos}
\usepackage{multirow}
\usepackage[skip=0pt]{caption}
\let\OLDthebibliography\thebibliography
\renewcommand\thebibliography[1]{
  \OLDthebibliography{#1}
  \setlength{\parskip}{1pt}
  \setlength{\itemsep}{1pt plus 0.3ex}
}

\title{Gravitational Wave Follow-Up Using Low Energy Neutrinos in IceCube DeepCore}
 \ShortTitle{IceCube DeepCore for Gravitational Wave Follow-up}

\author{The IceCube Collaboration \\{\normalsize \normalfont(a complete list of authors can be found at the end of the proceedings)}}

\emailAdd{aswathi.balagopalv@icecube.wisc.edu}

\abstract{The IceCube DeepCore is a dense infill array of the IceCube Neutrino Observatory at the South Pole. While IceCube is best suited for detecting neutrinos with energies of several 100 GeV and above, DeepCore allows to probe neutrinos with lower energies. We focus on a sample of neutrinos with energies above approximately 10 GeV, which was originally optimised for oscillation experiments. Recently, it has been adapted to enable searches for transient sources of astrophysical neutrinos in the sky. In particular, this low-energy dataset can be used to conduct follow-up searches of gravitational wave transients detected by the LIGO-Virgo instruments. A study of this, which complements IceCube’s follow-up of gravitational wave events using high-energy neutrino samples, will be discussed here.

\vspace{4mm}
{\bfseries Corresponding authors:}
Aswathi Balagopal V.$^{1*}$, Raamis Hussain$^{1}$, Alex Pizzuto$^{1}$\\
{$^{1}$ \itshape Wisconsin IceCube Particle Astrophysics Center, University of Wisconsin, Madison, WI, 53703, USA}\\[4mm]
$^*$ Presenter

\FullConference{37$^{\rm{th}}$ International Cosmic Ray Conference (ICRC 2021)\\
		July 12th -- 23rd, 2021\\
		Online -- Berlin, Germany}

}

\begin{document}
\maketitle
\section{Introduction}\label{sec:intro}
The IceCube Neutrino Observatory at the South Pole is a cubic-kilometer detector that observes neutrinos of both astrophysical and atmospheric origin. It consists of 86 strings drilled deep into the ice shelf, with digital optical modules (DOMs) mounted on them. The horizontal spacing of these strings is $125\,\mathrm{m}$ and the vertical spacing of the optical modules is $17\,\mathrm{m}$ \cite{Achterberg:2006md}.

IceCube has an infill array, called the IceCube DeepCore, that consists of 8 strings in the center of the main array, with optical modules optimized for the detection of neutrinos with energies lower than those detected by the main array \cite{Collaboration:2011ym}. This is achieved by reducing the horizontal string spacing to $72\,\mathrm{m}$ and the vertical spacing between the DOMs to $7\,\mathrm{m}$. The DOMs mounted on the DeepCore strings also have a higher quantum efficiency. IceCube DeepCore, with its special configuration, can therefore lower the detection threshold of neutrinos down to $\sim$ $10\,\mathrm{GeV}$, while IceCube's main array detects neutrinos with energies above 100s of GeV.

IceCube DeepCore is traditionally used for neutrino-oscillation studies. However, the lower energy threshold of DeepCore allows for a unique opportunity to probe for transient-astrophysical sources in the neutrino sky within the energy range of 10-100s of GeV. There exist several analyses within IceCube that utilise the DeepCore data for this purpose \cite{Abbasi:2020ddb, Larson:2021icrc, Aartsen_2016}. In this analysis, we focus on using such a dataset utilising DeepCore to search for low-energy neutrino counterparts to gravitational-wave (GW) events detected by the LIGO-Virgo detectors \cite{LIGOScientific:2018mvr}.

\section{Event Selection}
\begin{figure}[h]
\centering
\includegraphics[width=.8\textwidth]{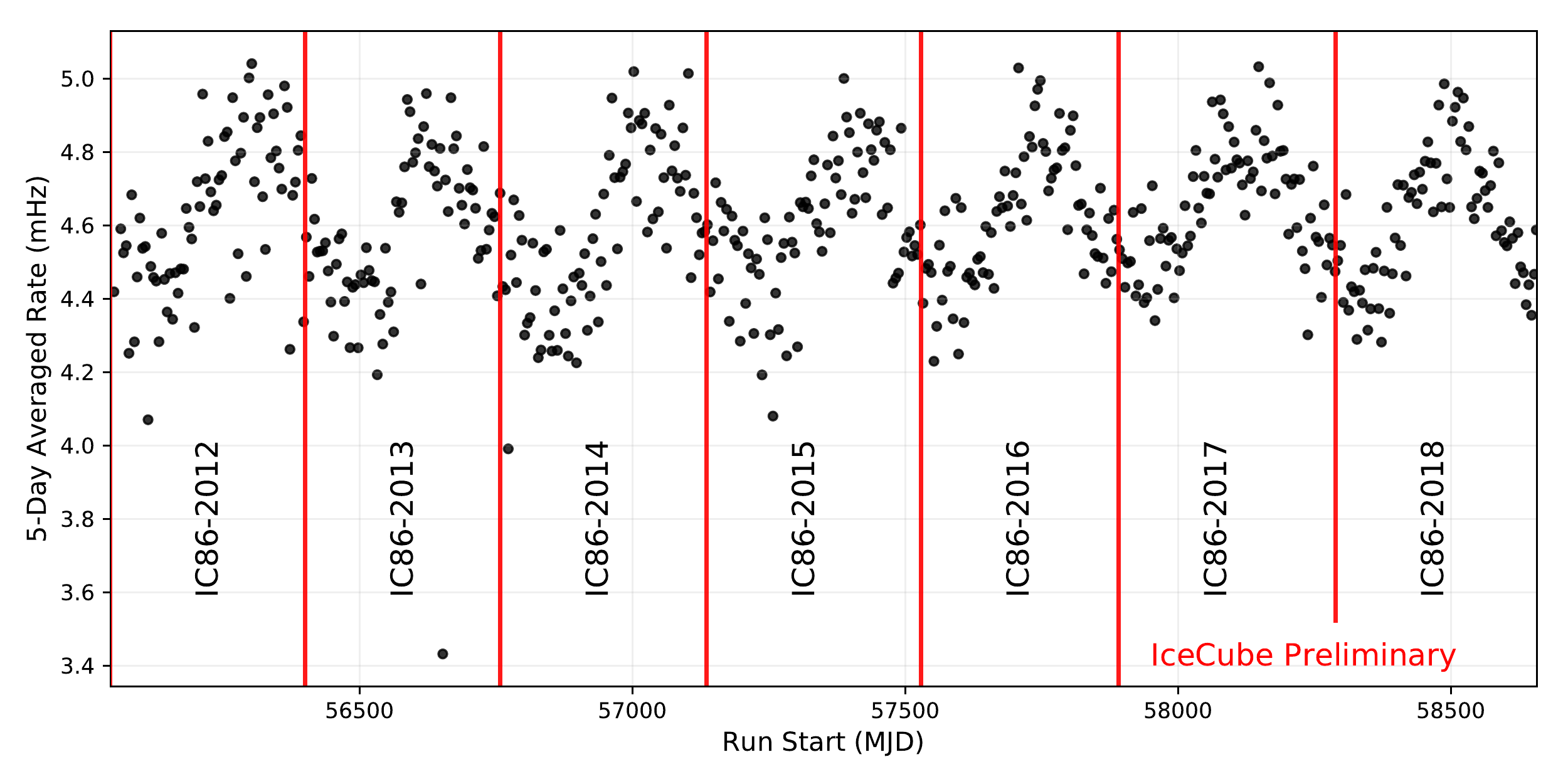}\caption{The 5-day averaged GRECO rate for the years 2012-2018. The averaged rate shown here has a sinusoidal behaviour due to the seasonal variations in the atmospheric neutrino flux \cite{Larson:2021icrc}.}\label{fig1}\end{figure}

An event selection has specifically been designed to select neutrinos with all flavours that interact inside the DeepCore array. This event selection, called GRECO (GeV Reconstructed Events with
Containment for Oscillations), was originally developed for tau-neutrino appearance studies, using boosted decision trees \cite{Aartsen:2019Larson}. The dataset, which starts from the year 2012, has an average rate of $4.5\,\mathrm{mHz}$ as shown in Figure \ref{fig1}. A stable rate like this allows us to search for transients within the dataset.

\begin{figure}[h]
\centering
\includegraphics[width=.8\textwidth]{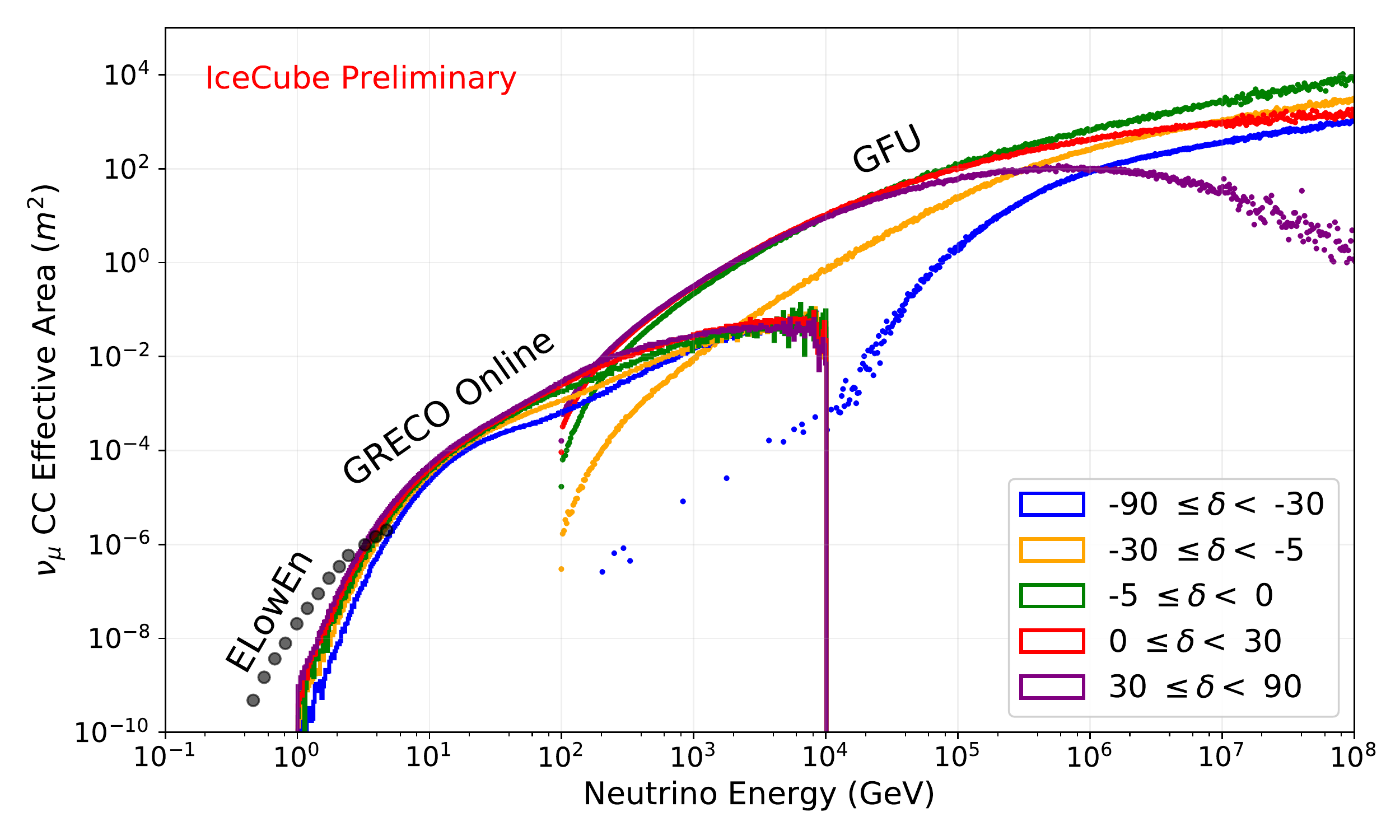}\caption{The effective area of the dataset using DeepCore compared to the high-energy tracks event selection in IceCube, and a low-energy event selection. The GRECO dataset, which is effective in the 10-100s of GeV range has nearly-uniform effective areas which varies by less than an order of magnitude across the whole sky \cite{Larson:2021icrc}.}\label{fig2}
\end{figure}

This event selection is applied to the whole sky, and therefore does not restrict the directions in which we can search for neutrinos. Figure \ref{fig2} shows the effective area\footnote{The area of an equivalent detector with 100\% efficiency for observing the incoming flux.} of the GRECO dataset for charged-current interactions from muon neutrinos in comparison to the high-energy tracks selection in IceCube (GFU: Gamma-ray Follow-Up \cite{GFU2016}) and a few-GeV event selection (ELOWEN \cite{DeWasseige:2019vN}). As shown in the figure, the GRECO dataset complements the effective areas of GFU and ELOWEN in the 10-100s of GeV energy range. Unlike the high-energy event sample, the GRECO dataset has comparable effective areas in both the Northern and Southern Hemispheres. This enhances the capabilities to probe for transient neutrinos across the whole sky.

The events within the GRECO dataset have worse angular resolutions compared to the higher-energy neutrinos observed by IceCube. In some cases. the angular uncertainty ($\sigma$) is as good as $1^\circ$, while in others, it is as bad as $70^\circ$. In the cases where the angular uncertainty is large the analysis essentially becomes a counting experiment. The GRECO sample used for this analysis has its angular uncertainties derived from a random forest algorithm which was trained on both charged-current and neutral-current interactions for neutrinos of all flavours \cite{Larson:2021icrc}.

\section{Analysis Method}
We use the GRECO dataset to search for neutrinos within the energy range of 10-100s of GeV that are potential counterparts to gravitational-wave events detected by the LIGO-Virgo instruments. Here, we focus on the 11 GW events observed during the O1 and O2 runs between 2015 and 2017 published in the GWTC-1 catalog of LIGO-Virgo \cite{LIGOScientific:2018mvr}.

The analysis method focuses on searching for neutrinos within a time window of $1000\,\mathrm{s}$ ($\pm 500\,\mathrm{s}$) centered around the time of the gravitational-wave event. These are the signal neutrinos that we are interested in, and are expected to have a spatial and temporal correlation with the GW events. Any other observed neutrino will be considered as background. The procedure is similar to the GW follow-up procedure used by the analysis based on the high-energy neutrino selection of IceCube (GFU) \cite{Aartsen:2020mla, Hussain:2021icrc}.

The main ingredient in the analysis is the unbinned likelihood that is used to search for transient sources of astrophysical  neutrinos, with $n_{s}$ signal neutrinos having a spectral index given by $\gamma$,
\begin{equation}
 \mathcal{L} = \frac{(n_{s}+n_{b})^{N}}{N!}e^{-(n_{s}+n_{b})}\,\prod_{i=1}^{N} \left(\frac{n_s\mathcal{S}_{i}}{n_{s}+n_{b}} + \frac{n_b \mathcal{B}_{i}}{n_{s}+n_{b}}\right).
\end{equation}
The first term represents the poisson probability of observing $N$ events. Here $n_{b}$ is the expected number of background events. The second term in the likelihood is a product over the probabilities of each event $i$ and includes a term for the signal and background probabilities of each event, given its energy and direction. $\mathcal{S}_{i}$ is the probability density function (PDF) for the signal and $\mathcal{B}_{i}$ is that for the background. The signal and background PDFs each consist of spatial and energy-dependent terms.

This likelihood is used to evaluate the test statistic (TS) to conduct the hypothesis test. A background-only null hypothesis (with $n_s\,=\,0$) is compared to a signal hypothesis with $n_s$ signal events and a spectral index given by $\gamma$. This test statistic is given by

\begin{equation}
 \mathrm{Test\,Statistic\,(TS)} = \mathrm{max.} \left\{\, 2\,\mathrm{ln}\left( \frac{\mathcal{L}_k(n_{s},\,\gamma)\,.\,w_k}{\mathcal{L}_k(n_s\,=\,0)} \right)\,\right\},
\end{equation}
where $k$ runs over each pixel in the sky. The calculation contains a spatial-prior term ($w_k$) in the numerator \cite{refId0}. This spatial prior is derived from the healpix skymap of the GW event, which divides the sky into 49152 equally-sized pixels. The spatial prior is a normalised penalty at each pixel which depends on the GW probability at that pixel in the sky. The penalty term $2\,\mathrm{ln} (w_k)$ has a maximum value of zero (which falls on the pixel with the maximum probability) and transitions to negative values (as the probability decreases). The maximum likelihood with respect to $n_s$ and $\gamma$ is estimated at each pixel in the sky, from which the TS value is obtained. This is performed within a time window of $\pm\,500\,\mathrm{s}$ with respect to the time of the GW event. 

\begin{figure}[h]

\includegraphics[width=.549\textwidth]{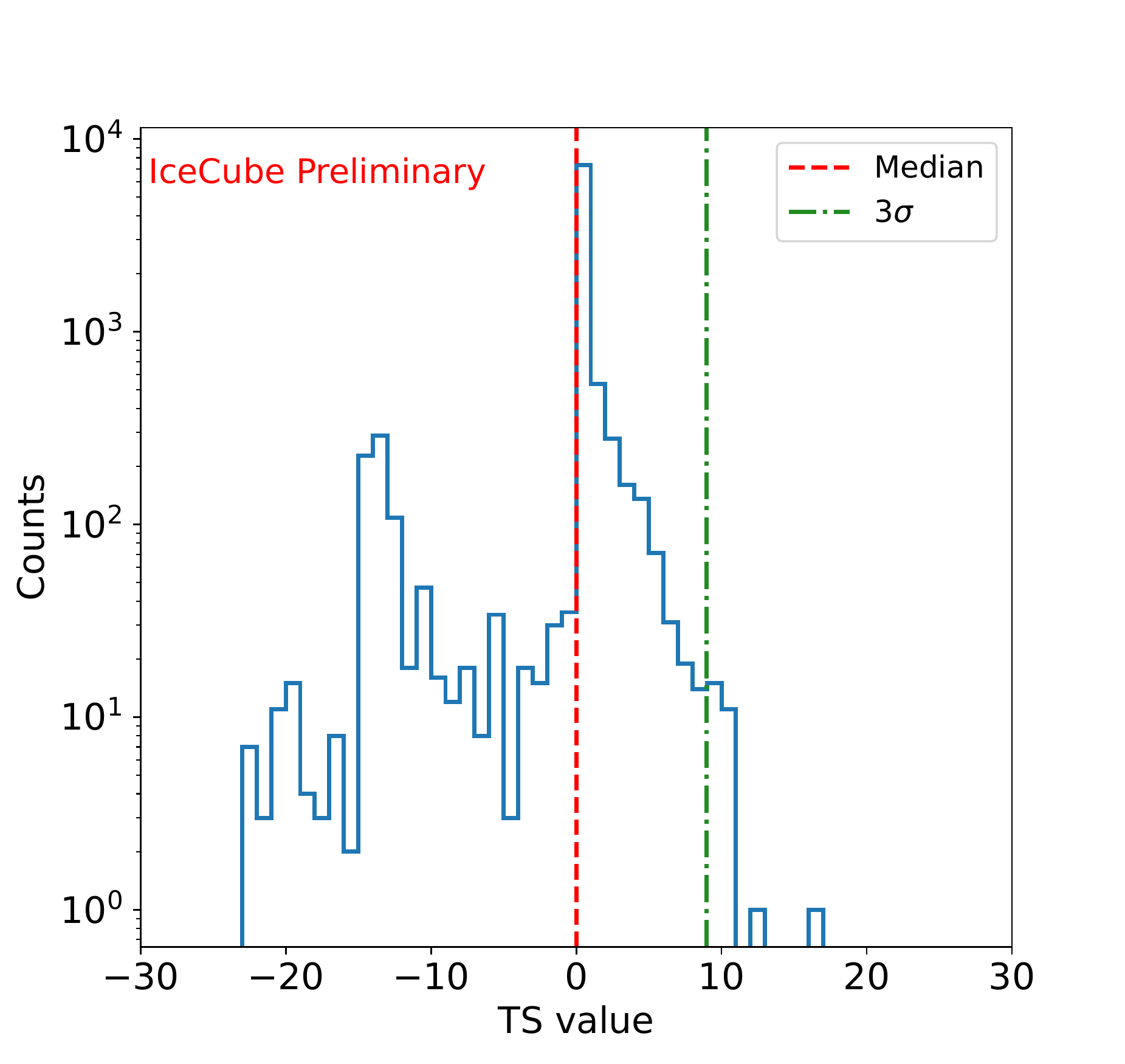}
\includegraphics[width=.5\textwidth]{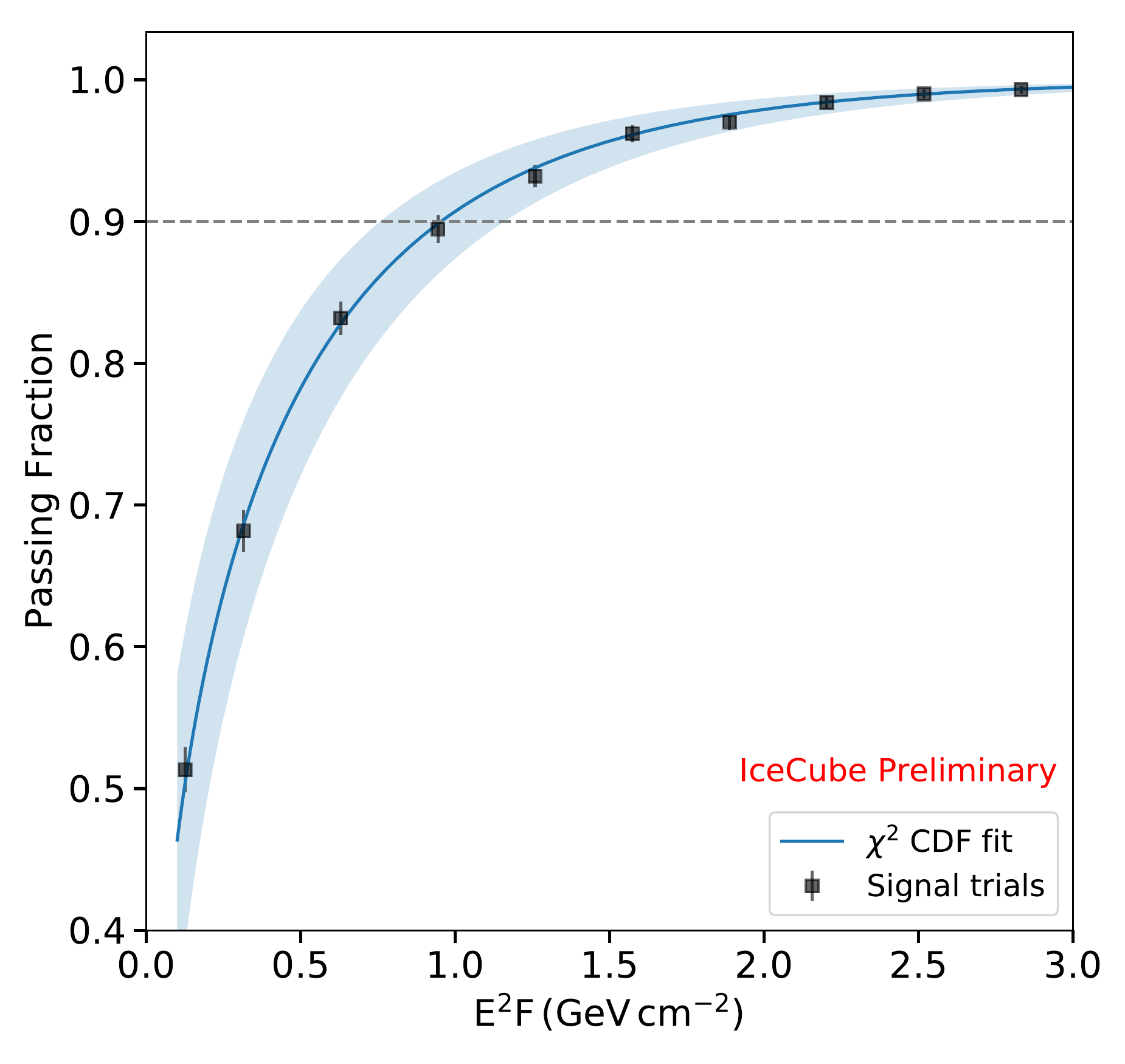}
\caption{Left: The background TS distribution (with spatial prior) for an example gravitational wave event. Right: Passing fractions for determining the 90\% sensitivity for GW170608. Here, $F$ is the time-integrated flux (within $1000\,\mathrm{s}$) given by $F\,=\,\frac{\mathrm{d}N}{\mathrm{d}E\mathrm{d}A\mathrm{d}t}\cdot \Delta t\,=\,\phi_{0}\cdot\left(\frac{E}{E_{0}}\right)^{-2}\,\mathrm{[GeV^{-1}\,cm^{-2}]}$}\label{fig3}
\end{figure}

We first construct a background distribution of the TS values using 10,000 pseudo experiments, for each GW event. This distribution for an example event (here GW170818) is shown in Figure \ref{fig3}. The background distribution is constructed from data by randomly sampling their arrival times from 5 days of data within the time of the GW event, while keeping their direction preserved in detector coordinates. This keeps the time structure and the declination dependence of the data, while altering the right ascension. 

There are certain features to the background TS distribution which becomes evident in Figure \ref{fig3}.
Several of the TS values are negative in the distribution. These arise from neutrino events that fall on the sky where they are penalized heavily by a negative-spatial prior. 
Apart from this, it can also be seen that the background TS distribution has certain structures to it. These structures correspond to the structures seen in the gravitational wave skymap, when observed in its log space. That is, when looking at the skymap in terms of $2\,\ln(w_k)$. 

\section{Analysis Performance}
We compute the sensitivity\footnote{The sensitivity of the analysis is defined as the median expected 90\% CL upper limit on the flux normalization in case of pure background. The discovery potential is defined as the signal strength that leads to a $3\sigma$ deviation from background in 90\% of the cases \cite{Aartsen:2018ywr}.} of the GRECO dataset to each gravitational-wave event by adding artificial neutrino signals from a Monte-Carlo dataset with a spectral index of $\gamma\,=\,2$ into a background-like dataset. The TS value obtained from such a pseudo experiment is compared to the median of the background TS distribution. This is repeated for a set of pseudo experiments for the same injected flux. The passing fraction, defined as the number of trials where the injected neutrino flux has a TS value larger than the median TS value of the background distribution, is determined for each flux level. The value of the flux at which the passing fraction becomes 0.9 is the 90\% sensitivity to the neutrino flux corresponding to a given GW event. This is illustrated in Figure \ref{fig3} for an example event GW170608. The per-flavour flux sensitivity is determined by fitting a $\chi^2$ cumulative distribution function (CDF) to the passing-fraction curve. The 90\% discovery potentials are calculated in the same manner as the sensitivity, but by comparing the TS values to the $3\sigma$ value obtained from the background TS distribution.

This procedure is performed for all 11 GW events detected by LIGO-Virgo during the O1 and O2 runs. In this work we show the sensitivities for a flux of muon neutrinos only, and the sensitivities with neutrinos of all-flavours is a work-in-progress. The obtained 90\% sensitivities and 90\% discovery potentials are shown in Table \ref{tab1} for each GW event. Also shown in the table are the corresponding number of neutrinos ($\mathrm{N_{inj}}$) that are needed to achieve the 90\% sensitivity and the 90\% discovery potential.

\begin{table}[h]
\centering
\captionsetup{justification=centering}
\caption{90\% sensitivities and 90\% discovery potentials (for $\nu_\mu+\bar{\nu}_\mu$ alone) to the GW events in the O1 and O2 runs.}\label{tab1}
\begin{tabular}{|c|c|c|c|c|}
\hline
\multirow{2}{*}{GW} & \multicolumn{2}{c|}{Sensitivity}              
& \multicolumn{2}{c|}{Discovery Potential}       \\ \cline{2-5} 
                    & \begin{tabular}[c]{@{}c@{}} $E^2F$ (GeV cm$^{-2}$)\end{tabular} & \begin{tabular}[c]{@{}c@{}}$\mathrm{N_{inj}}$\end{tabular} & \begin{tabular}[c]{@{}c@{}}$E^2F$ (GeV cm$^{-2}$)\end{tabular} & \begin{tabular}[c]{@{}c@{}}$\mathrm{N_{inj}}$\end{tabular} \\ \hline
GW150914  & 5.74  & 3.16   & 13.87    & 7.64 
\\ \hline
GW151012  & 2.24  & 3.23  & 13.8  & 19.73 
\\ \hline
GW151226   & 2.5   & 3.48  & 13.61  & 18.78      \\ \hline
GW170104 & 1.44   & 2.68   & 6.69 & 12.48       \\ \hline
GW170608  & 0.96  & 2.12    & 2.28  & 4.98 
\\ \hline
GW170729  & 5.75   & 4.52  & 15.42 & 11.91      \\ \hline
GW170809  & 4.62 & 2.76  & 12.73  & 7.58        \\ \hline
GW170814  & 5.88  & 3.08  & 14.65 & 7.08        \\ \hline
GW170817 & 5.00  & 2.86  & 11.51  & 6.69         \\ \hline
GW170818  & 1.07 & 2.24  & 1.94 & 4.07          \\ \hline
GW170823 & 2.85 & 3.58 & 12.59  & 16.01         \\ \hline
\end{tabular}
\end{table}

\begin{figure}[h]
\centering
\includegraphics[width=.65\textwidth]{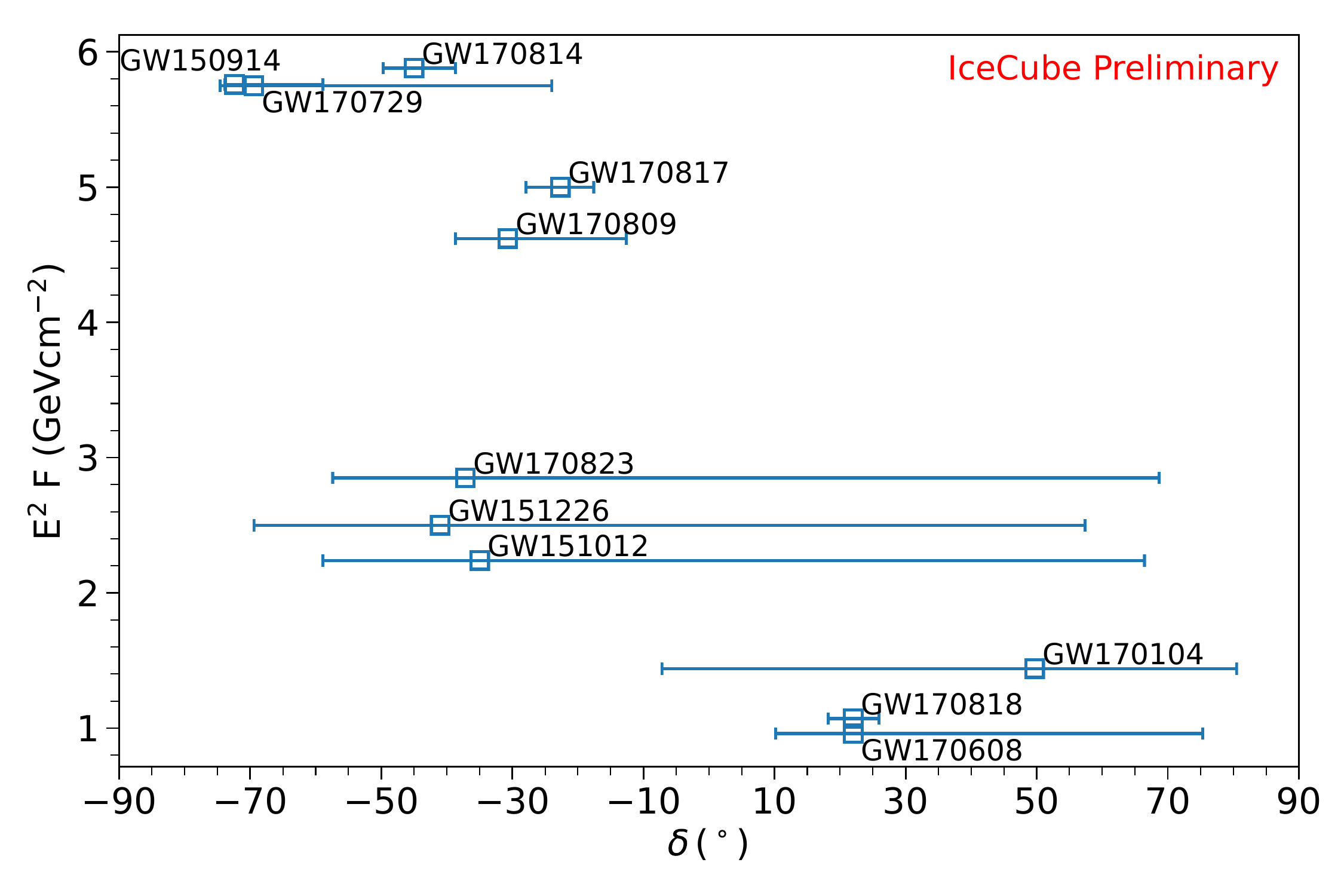}
\caption{Sensitivities ($\nu_\mu+\bar{\nu}_\mu$) of all events in the O1 and O2 runs. The markers show the declination of the maximum-probability pixel in the GW skymap and the error bars shows the 68\% probability region in the skymap.}\label{fig4}
\end{figure}

Figure \ref{fig4} shows the sensitivities of all the 11 GW events in the O1 and O2 runs together on the $y$-axis, compared to their respective declinations on the $x$-axis. The markers represent the declination of the maximum-probability pixel in the GW skymap, and the error bars represent the declinations covered by the region in the skymap that contains 68\% of the probabilities. From the figure, a declination-dependent behaviour is clear for the sensitivities. In general, the sensitivities are better in the Northern Hemisphere than in the Southern Hemisphere. However, this difference is within the same order of magnitude. This difference can be accounted for by the slightly lower effective area seen in the Southern Hemisphere than in the Northern Hemisphere for the GRECO dataset. Another contribution to the varying flux sensitivities can come from the slightly larger background rate in the Southern Hemisphere.

\begin{figure}[h]
\centering
\includegraphics[width=.7\textwidth]{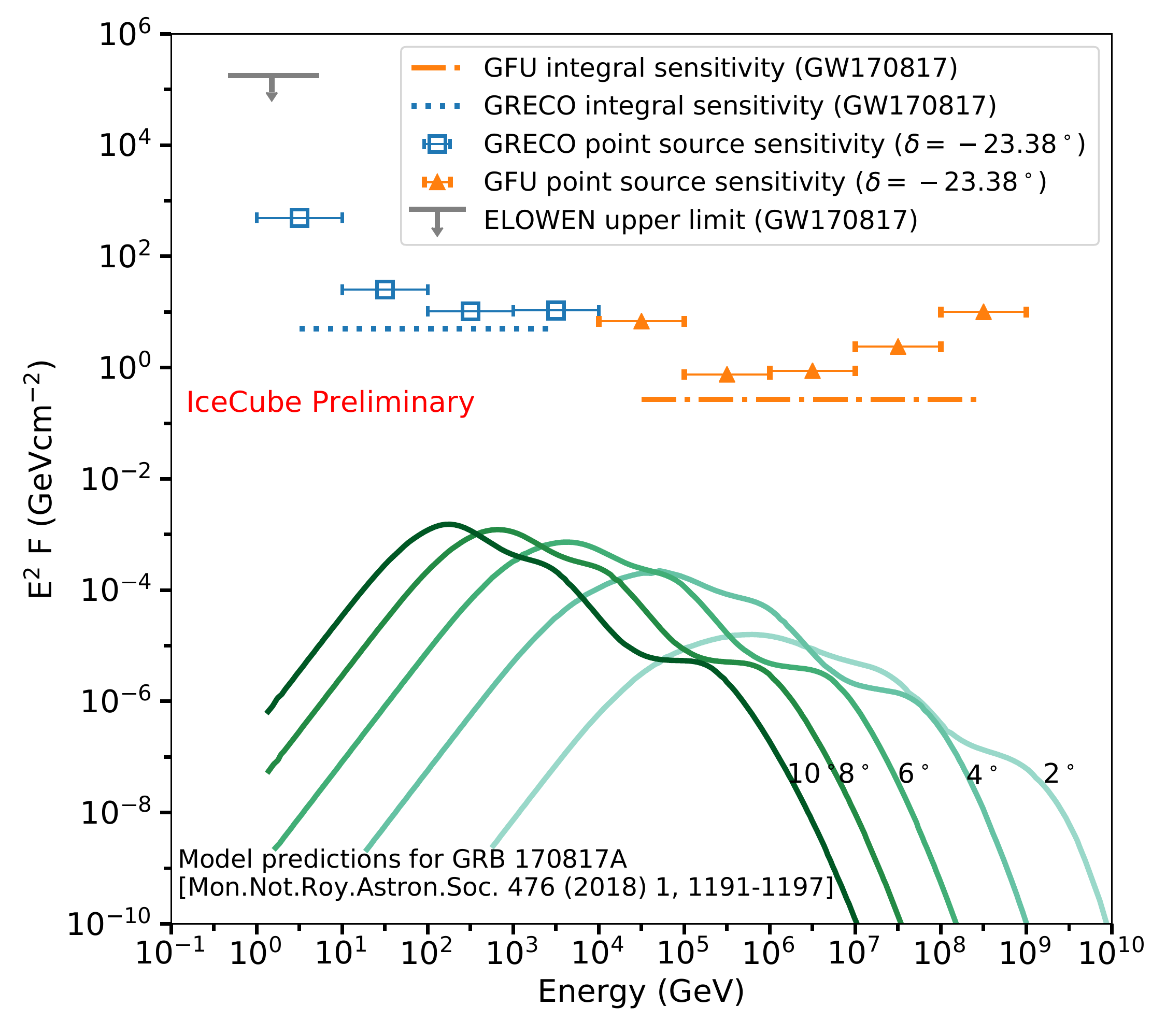}
\caption{The scope of various GW follow-up analyses within IceCube. Differential sensitivities (point source) of the GRECO and GFU samples, integral sensitivities of the GRECO and GFU samples to GW170817 \cite{Aartsen:2020mla} and the upper limit to GW170817 from ELOWEN \cite{DeWasseige:2019vN} are shown. All GRECO sensitivities shown here are for muon neutrinos only. Model predictions for GW170817 are also shown for different observation angles \cite{10.1093/mnras/sty285}.}\label{fig5}
\end{figure}

Figure \ref{fig5} shows a comparison of the sensitivities of various GW follow-up analyses of IceCube within the $1000\,\mathrm{s}$ time window which is used in this analysis. The integral sensitivity of the GRECO sample to the neutrino flux correlated to GW170817 is represented by the dotted line in the figure. The differential point source sensitivity of the GRECO dataset in the direction $\delta\,=\,-23.38^\circ$ (which is the direction of the host galaxy, NGC 4993, corresponding to the event GW170817 \cite{Abbott_2017}) is also shown in the figure (open squares). This is calculated by considering an $E^{-2}$ spectrum within each decade of energy. It is observed that the point-source sensitivity is nearly the same as the sensitivity with the spatial prior applied, for the GRECO dataset. This is particularly true for the cases where the GW skymaps are well localised. For events with extended GW skymaps these sensitivities are within the same order of magnitude. The sensitivities (point source and with spatial prior) at other declinations are also comparable to that shown in this figure for the GRECO dataset. Figure \ref{fig5} also shows the differential point-source sensitivity of the GFU sample in the direction $\delta\,=\,-23.38^\circ$, derived in the same manner as that of the GRECO dataset. The integral sensitivity (with spatial prior) of the high-energy GFU sample to GW170817 \cite{Aartsen:2020mla} and the upper limit for GW170817 obtained from the ELOWEN sample \cite{DeWasseige:2019vN} are also shown. Although the GFU sample demonstrates the best sensitivities to GW events, as shown in the figure, the capability of the GRECO dataset to probe lower energies opens up a new window in which we can search for neutrino counterparts to GW events.

The figure also shows model predictions for a short GRB like 170817A presented in \cite{10.1093/mnras/sty285}. The predictions depict off-axis emission for a fixed assumption of the Lorentz factor ($\Gamma\,=\,30$) and baryonic loading ($\xi\,=\,100$). The various curves represent different observation angles ($2^\circ$, $4^\circ$, $6^\circ$, $8^\circ$ and $10^\circ$), where the observation angle is the angle between the edge of the jet and the observation axis. Only the curves corresponding to the emission within the sub-photospheric region are shown in Figure \ref{fig5}. The model predicts that the further the observer is from the jet axis, the more peaked the emission becomes at energies lower than $1\,\mathrm{TeV}$. Although the model in \cite{10.1093/mnras/sty285} is specifically for the case of GW170817, this is relevant for other gravitational wave events also, since off-axis observations are more likely than on-axis observations.
\section{Summary and Outlook}
IceCube has conducted follow-up searches to gravitational wave events at energies above several hundreds of GeV and at energies within a few GeV. This analysis focuses on conducting follow-up searches in the energy range of 10-100s of GeV using a dataset derived on events contained within IceCube DeepCore. In this analysis, we presented the sensitivities to the 11 events observed during the O1 and O2 runs of LIGO-Virgo. These sensitivities are for muon neutrinos alone and those for neutrinos of all flavours is currently being studied. A comparison of the sensitivity of this analysis with other analyses covering different energy ranges with IceCube shows complementarity between the analyses, effectively expanding the energies in which such studies can be conducted with IceCube. It can be anticipated that the sensitivities shown here can be improved with future analyses that use improved angular reconstruction for the GRECO event sample, for example by using neural networks. The future upgrade to the IceCube detector is also expected to have better sensitivities to GeV energy neutrinos originating from compact-binary mergers.

\bibliographystyle{ICRC}
\bibliography{references}



\clearpage
\section*{Full Author List: IceCube Collaboration}



\scriptsize
\noindent
R. Abbasi$^{17}$,
M. Ackermann$^{59}$,
J. Adams$^{18}$,
J. A. Aguilar$^{12}$,
M. Ahlers$^{22}$,
M. Ahrens$^{50}$,
C. Alispach$^{28}$,
A. A. Alves Jr.$^{31}$,
N. M. Amin$^{42}$,
R. An$^{14}$
% ...
\\

\noindent
% ...
$^{12}$ Universit{\'e} Libre de Bruxelles, Science Faculty CP230, B-1050 Brussels, Belgium \\
% ...
$^{14}$ Department of Physics and Laboratory for Particle Physics and Cosmology, Harvard University, Cambridge, MA 02138, USA \\
% ...
$^{17}$ Department of Physics, Loyola University Chicago, Chicago, IL 60660, USA \\
$^{18}$ Dept. of Physics and Astronomy, University of Canterbury, Private Bag 4800, Christchurch, New Zealand \\
% ...
$^{22}$ Niels Bohr Institute, University of Copenhagen, DK-2100 Copenhagen, Denmark \\
% ...
$^{28}$ D{\'e}partement de physique nucl{\'e}aire et corpusculaire, Universit{\'e} de Gen{\`e}ve, CH-1211 Gen{\`e}ve, Switzerland \\
% ...
$^{31}$ Karlsruhe Institute of Technology, Institute for Astroparticle Physics, D-76021 Karlsruhe, Germany  \\
% ...
$^{42}$ Bartol Research Institute and Dept. of Physics and Astronomy, University of Delaware, Newark, DE 19716, USA \\
% ...
$^{50}$ Oskar Klein Centre and Dept. of Physics, Stockholm University, SE-10691 Stockholm, Sweden \\
% ...
$^{59}$ DESY, D-15738 Zeuthen, Germany \\
% ...

\subsection*{Acknowledgements}

\noindent
USA {\textendash} U.S. National Science Foundation-Office of Polar Programs,
U.S. National Science Foundation-Physics Division,
U.S. National Science Foundation-EPSCoR,
Wisconsin Alumni Research Foundation,
Center for High Throughput Computing (CHTC) at the University of Wisconsin{\textendash}Madison,
Open Science Grid (OSG),
Extreme Science and Engineering Discovery Environment (XSEDE),
Frontera computing project at the Texas Advanced Computing Center,
U.S. Department of Energy-National Energy Research Scientific Computing Center,
Particle astrophysics research computing center at the University of Maryland,
Institute for Cyber-Enabled Research at Michigan State University,
and Astroparticle physics computational facility at Marquette University;
Belgium {\textendash} Funds for Scientific Research (FRS-FNRS and FWO),
FWO Odysseus and Big Science programmes,
and Belgian Federal Science Policy Office (Belspo);
Germany {\textendash} Bundesministerium f{\"u}r Bildung und Forschung (BMBF),
Deutsche Forschungsgemeinschaft (DFG),
Helmholtz Alliance for Astroparticle Physics (HAP),
Initiative and Networking Fund of the Helmholtz Association,
Deutsches Elektronen Synchrotron (DESY),
and High Performance Computing cluster of the RWTH Aachen;
Sweden {\textendash} Swedish Research Council,
Swedish Polar Research Secretariat,
Swedish National Infrastructure for Computing (SNIC),
and Knut and Alice Wallenberg Foundation;
Australia {\textendash} Australian Research Council;
Canada {\textendash} Natural Sciences and Engineering Research Council of Canada,
Calcul Qu{\'e}bec, Compute Ontario, Canada Foundation for Innovation, WestGrid, and Compute Canada;
Denmark {\textendash} Villum Fonden and Carlsberg Foundation;
New Zealand {\textendash} Marsden Fund;
Japan {\textendash} Japan Society for Promotion of Science (JSPS)
and Institute for Global Prominent Research (IGPR) of Chiba University;
Korea {\textendash} National Research Foundation of Korea (NRF);
Switzerland {\textendash} Swiss National Science Foundation (SNSF);
United Kingdom {\textendash} Department of Physics, University of Oxford.
\end{document}